\documentclass{myelsart}

\usepackage{graphics}
\usepackage[T1]{fontenc}
\usepackage{array}
\usepackage{colordvi}

\usepackage{amssymb}
\begin{document}
\begin{frontmatter}


\title{A quantitative model for presynaptic free $\rm Ca^{2+}$ dynamics during
different stimulation protocols}


\author{Frido Erler\corauthref{cor1}}\ead{erler@theory.phy.tu-dresden.de},
\author{Michael Meyer-Hermann},
\author{Gerhard Soff}


\corauth[cor1]{Corresponding Author. Tel.: +49 351 46333841; Fax: +49 351 46337299.}


\address{Institut für Theoretische Physik, Technische Universität Dresden,
01062 Dresden, Germany}

\begin{abstract}
The presynaptic free $\rm Ca^{2+}$ dynamics of neurons due to various stimulation
protocols is investigated in a mathematical model. The model includes $\rm Ca^{2+}$ 
influx through voltage-dependent
$\rm Ca^{2+}$ channels, $\rm Ca^{2+}$ buffering by endogenous and exogenous buffers
as well as $\rm Ca^{2+}$ efflux through ATP-driven plasma membrane $\rm Ca^{2+}$
pumps and  $\rm Na^{+}$/$\rm Ca^{2+}$ exchangers. We want to support a specific way of
modeling which starts on the level of single proteins. Each
protein is represented by characteristics that are determined by single protein 
experiments and that are considered to be widely independent of neuron 
types. This allows the applications of the model to different classes of 
neurons and experiments. The procedure is demonstrated for single 
boutons of pyramidal neurons of the rat neocortex. The corresponding
fluorescence measurements of Koester \& Sakmann 
(2000, J. Physiol., 529, 625) are quantitatively reproduced.
The model enables us to reconstruct the free $\rm Ca^{2+}$ dynamics in neurons as it would
have been without fluorescence indicators starting from the fluorescence
data. We discuss the different $\rm Ca^{2+}$ responses and find that during
high-frequency stimulation an accumulation of free $\rm Ca^{2+}$ occurs above some 
threshold stimulation frequency. The threshold frequency
depends on the amount of fluorescence indicator used in the experiment. 
\end{abstract}

\end{frontmatter}

\section{Introduction\label{chapter intro}}
$\rm Ca^{2+}$ is an important cellular messenger. In particular, $\rm Ca^{2+}$
plays a key role in synaptic transmission. For example, it triggers
the secretion of neurotransmitter \cite{Llinas_etal81,Katz69} and
is involved in synaptic plasticity \cite{Bliss_Collingridge93} being most likely
related to the neural bases of memory \cite{Tsien00}. Important phenomena in this
context are presynaptic facilitation, short-term potentiation (STP), 
long-term potentiation (LTP) and long-term depression (LTD). It has been shown that the
induction of long-term effects (LTE) are related to a prior $\rm Ca^{2+}$ influx into
the neuron \cite{Ltebooks}.

There exist several theoretical and numerical approaches
to describe intracellular calcium dynamics. 
Some models investigate
presynaptic $\rm Ca^{2+}$ dynamics in response to single action potentials
\cite{Sinha_etal97,Helmchen_etal97,Sabatini_Regehr98}.
The role of free calcium in transmitter release and in
synaptic facilitation has been discussed 
in detail \cite{Sinha_etal97,Fogelson_Zucker85,Rozov_etal01,Cooper_etal96}.
These models focus on the geometry of the presynaptic compartment and the
distribution of calcium transporting proteins in the membrane.
The role of intracellular organelles with the capability
to store and release free calcium (e.g. endoplasmatic and sarcoplasmatic
reticula) has been analyzed \cite{Keizer_etal98,Schiegg_etal95,DeSchutter_Smolen98}.
Also the role of slow and fast intracellular buffers has been
investigated \cite{Smith_etal96a,Matveev_etal02}. 
Indeed, most of the incoming calcium ions
are bound to some intracellular buffers 
\cite{Neher_Augustine92}, such that this issue is very important.

The philosophy of the above models is to focus on the spatial
distribution of calcium ions in different compartments of the cell.
The connection of calcium and transmitter release has been
investigated mostly in two giant synapses: 
the calyx of Held \cite{Meinrenken_etal02}
and the squid giant synapse \cite{Llinas99}.
For example it has been discussed for the squid giant synapse
that the calcium enhancement
in the direct neighborhood of the channels is related to
transmitter release and facilitation \cite{Zucker_Fogelson86}. 
To this end the channels
enter this model as discrete entities. We would like to reconsider
this approach on the basis of new data that are available today.

Single channel properties have become more and more available
\cite{Sakmann95}. In the time of the work of Fogelson and Zucker
the exact single protein currents and open dynamics 
were not available. Therefore, these data were modeled with
some free parameters. The question we would like to address 
is whether we have sufficient information on single protein 
properties at hand in order to explicitly and quantitatively
describe the activation and activity of each protein involved
in the transmembrane calcium flux.

To this end we will proceed as follows:
\begin{itemize}
\item
Instead of describing large squid giant synapse we restrict
ourselves in a first step 
to rather small synapses. We focus
on the average calcium transient in the synapse and 
neglect the effect of calcium diffusion. 
This prevents
us from considering LTE or synaptic facilitation in
detail due to the lack of detailed spatial information
on inhomogeneous calcium distributions. 
The average calcium
transient is nevertheless an important indicator
since the exact amplitude and time course of calcium,
i.e. the exact shape of the calcium transient, is believed
to be important for transmitter release \cite{Barrett_Stevens72}.
The average calcium transient is also suitable for the analysis
of the quality of a model based on single protein properties.
\item
The calcium dynamics is described by well known
ordinary differential equations. Within these equations
the single protein properties are isolated and treated
separately. 
\item
The model parameters concerning the behaviour of single
proteins are determined on the basis of independent 
single protein experiments. This set of parameters will
be called {\it universal} because it will be assumed that
the properties of the proteins 
do not primarily depend on the neuron under consideration.
This is surely an approximation which has to be further
validated.
\item
The model parameters concerning the specific setup of
an experiment, i.e. extension of the synaptic compartment,
calcium concentrations, stimulation protocol, use of
fluorescence indicator, etc. are adjusted to a specific
experiment measuring calcium transients in response to
various stimulation protocols in single 
boutons of pyramidal neurons of the rat neocortex
\cite{Koester_Sakmann00}. These {\it specific} parameters
have to be adjusted for each experiment separately.
\item
The remaining parameters (in our case this will be the
protein densities in the membrane of the neuron) are used
in order to fit the data. It is important that the number
of fit parameters is minimized in order to get a suitable
test of the model.
\end{itemize}

Note, that the separation of universal and specific parameters
in the model opens the possibility to use the universal
parameters, which have been determined once in independent
experiments, for different experimental setups. 
On one hand, this considerably
reduces the number of fit parameters and strongly restricts
possible outcomes of the model. On the other hand,
this allows to compare the results of different experiments
using different protocols or calcium indicators.
Indeed, it has been mentioned before, that the amount and
type of calcium indicator changes the calcium transients
in a non-negligible way \cite{Smith_etal96a,Matveev_etal02}. 
Therefore, we consider this model approach as a first step
towards a link between different experiments opening the
possibility of a comparative approach to calcium data.

Our main interest is a quantitative description of presynaptic free
$\rm Ca^{2+}$ dynamics in response to single action potentials
and to different stimulation protocols. We apply the model to single 
boutons of the pyramidal neurons in the rat neocortex and compare 
the results to the corresponding experiment \cite{Koester_Sakmann00}. 
This experiment provides most of the necessary informations
in order to determine the specific parameters of the model.
We will look for characteristics in $\rm Ca^{2+}$ time courses
which are specific for the used stimulation protocols.
Furthermore, we will analyze the influence of the calcium indicator
on the results of the experiment and discuss the possibility to reconstruct
the calcium transient as it would have been without indicator.

\section{Methods\label{chapter model}}

The intracellular free $\rm Ca^{2+}$ dynamics due to a predefined, time
dependent membrane potential of a presynaptic neural domain is described
by using a set of ordinary differential equations. This neural domain
is represented by a cell membrane which separates the extracellular
space from the intracellular space. Specific membrane proteins are
embedded into this membrane. Some of these membrane proteins and in
addition some intracellular proteins determine the dynamics of the
free $\rm Ca^{2+}$ ions concentration \cite{Llinas82,Carafoli02,Magee_etal98}.

The model geometry consists of a homogenous reaction volume (one-compart\-ment)
which is situated in a homogeneous extracellular space. We are using
a spherical symmetry of the reaction volume. However, the results
remain valid for other shapes provided that the surface to volume
ratio is conserved. In general presynaptic boutons are relatively
small, i.e. having a diameter of the order of microns \cite{Koester_Sakmann00}.
In this study we assume that the process of diffusion of
$\rm Ca^{2+}$ ions is fast enough to reach an equilibrium distribution on
a time scale of one \( \rm {ms} \). This assumption has to be reviewed
considering larger reaction volumes as found in postsynapses. Consequently,
also the surface densities of membrane proteins have to be interpreted
as average values over the whole reaction surface.

Starting from a cell in equilibrium the application of a depolarizing
membrane potential \( V \) induces a free $\rm Ca^{2+}$ ion current between
the reaction volume and the extracellular space through diverse membrane
proteins. This leads to a change of the intracellular free $\rm Ca^{2+}$
concentration $c=\left[ \rm {Ca}_{\rm {free}}^{2+}\right]$ which is represented 
by an ordinary differential equation

\begin{equation}
\label{eq meaneq}
\frac{d}{dt}c=\frac{G}{\rm {zF}}\left\{ J_{\rm {i}}\left( V\left( t\right) ,
\, \overline{V}\left( c\right) \right) -J_{\rm {e}}\left( c\right) +L\right\}
+B_{\rm {en}}\left( c\right) +B_{\rm {ex}}\left( c\right)
\end{equation}

in analogy to other studies (see for example \cite{Smith_etal96b} eq.(1) or
\cite{Schiegg_etal95} eq.(9)). Every source and every sink of the free 
$\rm Ca^{2+}$ ions is determined by its own term. \( J_{\rm {i}} \) is the 
influx current density per membrane surface unit through voltage-dependent 
$\rm Ca^{2+}$ channels (VDCC) \cite{Tsien83,Bean89,Hess90}. It directly depends 
on the applied membrane potential \( V \) (the time dependence in Eq.~(\ref{eq meaneq})
is noted explicitly to emphasize that \( V\left( t\right)  \) is
treated as input signal), and the effective reversal potential \( \overline{V} \)
and is described in detail in section~\ref{chapter influx}. \( J_{\rm {e}} \)
is the efflux current density per membrane surface unit caused by
ATP-driven plasma membrane $\rm Ca^{2+}$ pumps (PMCA) 
\cite{Monteith_Roufogalis95,Penniston_Enyedi98,Carafoli92}
and $\rm Na^{+}$/$\rm Ca^{2+}$ exchangers (NCX) 
\cite{Blaustein_Lederer99,Strehler90,DiPolo_Beauge99}.
It depends on the free intracellular $\rm Ca^{2+}$ concentration \( c \)
and is described in detail in section~\ref{chapter efflux}. The
leakage surface current density \( L \) is determined by the steady state conditions 
and represents all currents which are not described in our model
in an explicit manner and that determine the equilibrium state
(section~\ref{chapter sss}). The terms \( B_{\rm {en}} \) and \( B_{\rm {ex}} \) 
stand for the action of endogenous and the exogenous buffers 
\cite{Miller95,Klee88,Pottorf_etal00}, respectively, which bind to and disengage 
from the free $\rm Ca^{2+}$ ions (see section \ref{chapter buffering}). The geometry
factor $G=O/V$ is the surface to volume ratio. It translates the current densities
(\( J_{\rm {i}} \), \( J_{\rm {e}} \), \( L \) ) into changes of
free intracellular $\rm Ca^{2+}$ concentration. \( \rm {z} \) and \( \rm {F} \)
are the valence charge of $\rm Ca^{2+}$ ions and the Faraday constant, respectively.

Note, that the separation of universal and specific parameters is
not reflected in Eq. (\ref{eq meaneq}). However, in the following sections the explicit
form of the currents entering the equation will be developed on the
level of single membrane proteins. It is at this point that the parameter
classification will be mirrored in the mathematical prescription.

\subsection{$\rm Ca^{2+}$ influx through VDCC\label{chapter influx}}

The VDCC is characterized by a current-voltage relation based on both
a voltage dependent current through the open pore \( I^{\rm {single}\, 
\rm {channel}}_{\rm {open}} \) as well as a voltage dependent opening probability 
\( g_{\rm {v}} \) \cite{Tsien83}. We use the following description of the $\rm
Ca^{2+}$ influx current density \( J_{\rm {i}} \) through VDCC \cite{Volfovsky_etal99}

\begin{equation}
\label{eq vdcc}
\begin{array}{cc}
J_{\rm {i}}\left( V\left( t\right) ,\, \overline{V}\left( c\right) \right) =
& \rho _{\rm {v}}\, \, g_{\rm {v}}\left( V\left( t\right) \right) \, \,
\underbrace{\overline{g}_{\rm {v}}\, \, \left( \overline{V}\left( c\right) -
V\left( t\right) \right) }\, \, .\\ & \, \, \, \, \, \, \, \, \, \, \, \, \,
\, \, \, \, \, \, \, \, \, \, \, \, \, \, \, \, \, \, \, I^{\rm {single}\,
\rm {channel}}_{\rm {open}}
\end{array}
\end{equation}

\( J_{\rm {i}} \) is weighted with the surface density of the channels
\( \rho _{\rm {v}} \) being the most important cell-type \emph{specific}
parameter. Practically all other parameters defining the single channel
properties belong to the class of universal parameters. The $\rm Ca^{2+}$
current through the open pore \( I^{\rm {single}\, \rm {channel}}_{\rm {open}} \)
is driven by the electrochemical gradient over the membrane \cite{Nernst89,Planck91}.
A current caused by changes of the membrane potential is followed
by changes of the intracellular $\rm Ca^{2+}$ concentration and these retroact
on the current. The $\rm Ca^{2+}$ current due to the membrane potential
gradient is approximated by a linear voltage-current relation with
open pore conductivity \( \overline{g}_{\rm {v}} \). This is justified
for physiologically relevant membrane potentials for which the VDCC
current-voltage relation has been found to be indeed nearly linear
\cite{Magee_Johnston95,Fisher_etal90,Griffith_etal94}. The important
potential difference entering Eq.~(\ref{eq vdcc}) is the one relative
to the $\rm Ca^{2+}$ reversal potential \( \overline{V} \) which dynamically
incorporates the $\rm Ca^{2+}$ concentration gradient over the membrane
into the model using Nernst equation \cite{Nernst88}

\begin{equation}
\label{eq reversal}
\overline{V}\left( c\right) =\frac{\rm {RT}}{\rm {zF}}\ln 
\left( \frac{c_{\rm {ext}}}{c}\right) -\Delta V_{\rm {eff}}\, \, .
\end{equation}

\( \rm {R} \) is the molar gas constant and \( \rm {T} \) is the
absolute temperature. \( c_{\rm {ext}} \) stands for the external
$\rm Ca^{2+}$ concentration. \( \Delta V_{\rm {eff}} \) corrects the exact
reversal potential for the linear approximation. As it has been found
that efflux $\rm Ca^{2+}$ currents through VDCC at voltages above the reversal
potential are negligible \cite{Hille92,Brown_etal81} 
we set \( I^{\rm {single}\, \rm {channel}}_{\rm {open}}=0 \)
for \( V>\overline{V} \). Independently from the definition of the reversal
potential by the Nernst equation it should be mentioned that its $\rm Ca^{2+}$
dependency has only minor effects on the $\rm Ca^{2+}$ dynamics. Note that the exact
value of \( \Delta V_{\rm {eff}} \) may depend on the ion concentrations in each cell. 

The time dependence of the single channel open probability \( g_{\rm {v}} \) is modeled 
by a single exponential approximation

\begin{equation}
\label{eq dyn openprob}
\frac{d}{dt}g_{\rm {v}}=\left( \widehat{g}_{\rm {v}}\left( V\right) -
g_{\rm {v}}\right) \frac{1}{\tau }
\end{equation}

 which may be related to experiment using data for the average number
of open channels in an ensemble of channels 
\cite{Magee_Johnston95,Griffith_etal94,O'Dell_Alger91}.
The open probability approaches its asymptotic value \( \widehat{g}_{\rm {v}}(V) \)
with a time constant \( \tau  \). \( \widehat{g}_{\rm {v}}(V) \)
is different for each membrane potential and is described by a sigmoidal
function

\begin{equation}
\label{eq asymptopprob}
\widehat{g}_{\rm {v}}\left( V\right) =\frac{1}{\exp \left( \left( V_{\rm {h}}-V\right) 
\frac{1}{\kappa }\right) +1}
\end{equation}

being the best approximation to experimental data 
\cite{Magee_Johnston95,Fisher_etal90,Griffith_etal94}.
Here, \( V_{\rm {h}} \) is the half activation voltage and \( \kappa  \)
describes the steepness of the asymptotic opening probability
\( \widehat{g}_{\rm {v}} \).

\subsection{$\rm Ca^{2+}$ buffering via endogenous and exogenous buffers
\label{chapter buffering}}

The main part of the intracellular $\rm Ca^{2+}$ is bound to an endogenous
buffer \cite{Neher_Augustine92,Belan_etal93,Belan_etal93b}. In addition, 
also a fluorescence indicator used in experiments acts as an exogenous buffer 
\cite{Helmchen_etal96}. In the rest state of the neuron the amount of free buffers 
is still large enough such that during stimulation the dominant part of incoming
free $\rm Ca^{2+}$ ions bind to these buffers \cite{Neher_Augustine92}.
The steadily ongoing binding and dissociation process of $\rm Ca^{2+}$ and
buffers strongly influence the resulting free $\rm Ca^{2+}$ dynamics. This
applies not only to the phase of growing $\rm Ca^{2+}$ concentration during
stimulation but also to the return of the $\rm Ca^{2+}$ concentration to
the base level. Therefore, we include both types of buffers explicitly
in the model using the following kinetic equations

\begin{equation}
\label{eq buffkin}
\frac{d}{dt}b_{\rm {en},\rm {ex}}=-k^{-}_{\rm {en},\rm {ex}}\, \, b_{\rm {en},\rm {ex}}
+k^{+}_{\rm {en},\rm {ex}}\, \, c\, \, \left( b^{0}_{\rm {en},\rm {ex}}-
b_{\rm {en},\rm {ex}}\right)
\end{equation}

for the endogenous and exogenous buffer 
\cite{Sabatini_Regehr98,Schiegg_etal95,Wagner_Keizer94}.
Here, \( b_{\rm {en}} \) and \( b_{\rm {ex}} \) are the concentrations
of bound buffers, \( b_{\rm {en}}^{0} \) and \( n_{\rm {ex}}^{0} \)
are the total concentrations of the intracellular buffer proteins.
\( k^{+}_{\rm {en},\rm {ex}} \) and \( k^{-}_{\rm {en},\rm {ex}} \)
are the rate constants in units of \( \frac{1}{\mu \rm {M}\, \rm {ms}} \)
and \( \frac{1}{\rm {ms}} \), respectively. As we do not take diffusion
into account, we neglect all spatial effects of mobile buffers. Also all effects
related to different types of buffers are neglected. Note, that 
the source and the sink terms for the buffers (Eq.~(\ref{eq buffkin})) are
the sink and the source terms, respectively, for the $\rm Ca^{2+}$ dynamics
(Eq.~(\ref{eq meaneq}))

\begin{equation}
\label{eq buffcalckin}
B_{\rm {en},\rm {ex}}\left( c\right) =-\frac{d}{dt}b_{\rm {en},\rm {ex}}\, \, .
\end{equation}

The rate constants are in general not available from experiment. We
use the realistic assumption that the buffer dynamics takes places
on short time scales compared to the $\rm Ca^{2+}$ dynamics, i.e. that the
buffer dynamics is always in a quasi steady state (rapid buffer
approximation \cite{Smith_etal96a,Neher_Augustine92,Neher98}). 
Note, that we neglect low affinity endogenous buffers in the
present model. The rapid buffer approximation has two important
consequences. First, on the short time scale the bound endogenous
and exogenous buffer concentrations \( b_{\rm {en},\rm {ex}} \) become
constant and can be written as

\begin{equation}
\label{eq buffqssresult}
b_{\rm {en},\rm {ex}}\left( c\right) =\frac{c\, \, b_{\rm {en},\rm {ex}}^{0}}
{K_{\rm {en},\rm {ex}}+c}\, \, .
\end{equation}

This implies that the bound buffer concentration is adapted instantaneously
to the free $\rm Ca^{2+}$ concentration at each time point \( t. \) Therefore,
we have

\begin{equation}
\label{eq buffimpl}
B^{\rm {qss}}_{\rm {en},\rm {ex}}\left( c\right) =-\frac{d}{dt}b_{\rm {en},\rm {ex}}
\left( c \right) = -\frac{d}{dc}b_{\rm {en},\rm {ex}}\left( c\right) \frac{d}{dt}c
\, \, .
\end{equation}

With the help of Eq.~(\ref{eq buffqssresult}) and Eq.~(\ref{eq buffimpl})
the $\rm Ca^{2+}$ dynamics (Eq.~(\ref{eq meaneq})) gets a more compact form:

\begin{equation}
\label{eq appmeaneq}
\frac{d}{dt}c=\frac{G}{\rm {zF}}\left\{ J_{\rm {i}}-J_{\rm {e}}+L\right\} 
\frac{1}{1+T_{\rm {en}}+T_{\rm {ex}}}\, \, .
\end{equation}

\( T_{\rm {en}} \) and \( T_{\rm {ex}} \) are dimensionless variables:

\begin{eqnarray}
T_{\rm {en}}\left( c\right)  & = & \frac{b^{0}_{\rm {en}}\, K_{\rm {en}}}
{\left( K_{\rm {en}}+c\right) ^{2}}\quad \rm{with}\quad K_{\rm {en}}=
\frac{k_{\rm {en}}^{-}}{k_{\rm {en}}^{+}}\nonumber \\
T_{\rm {ex}}\left( c\right)  & = & \frac{b_{\rm {ex}}^{0}K_{\rm {ex}}}
{\left( K_{\rm {ex}}+c\right) ^{2}}\quad \rm {with}\quad K_{\rm {ex}}=
\frac{k_{\rm {ex}}^{-}}{k_{\rm {ex}}^{+}}
\label{eq exopuffterm}
\end{eqnarray}

and \( \frac{1}{1+T_{\rm {en}}+T_{\rm {ex}}} \) is a correction factor
in Eq.~(\ref{eq appmeaneq}) which depends only on the free $\rm Ca^{2+}$
concentration. The great technical advantage of this approximation \cite{Wagner_Keizer94}
is that Eq.~(\ref{eq appmeaneq}) only involves the dissociation
constants \( K_{\rm {en},\rm {ex}} \) of the buffers instead of both
rate constants \( k_{\rm {en},\rm {ex}}^{+} \) and \( k^{-}_{\rm {en},\rm {ex}} \).
In general the dissociation constants are available from experiment.

\subsection{$\rm Ca^{2+}$ efflux through PMCA and NCX\label{chapter efflux}}

The PMCA and the NCX are proteins which actively transport $\rm Ca^{2+}$
ions through the membrane \cite{Carafoli91,Blaustein_Lederer99} and
therefore have to be described differently from pores. The PMCA pumps
$\rm Ca^{2+}$ ions against the electrochemical gradient using the energy
of ATP molecules. The NCX transports $\rm Ca^{2+}$ ions out of the neuron
by exchanging them with $\rm Na^{+}$ ions. The $\rm Ca^{2+}$ dependence of the
PMCA-kinetics observed in experiments is well described by the Hill
equation \cite{Elwess_etal97,Verma_etal96,Desrosiers_etal96}. The
activity of the PMCA and the NCX is limited to a maximum rate 
\cite{Caride_etal01,Blaustein_Lederer99,Strehler90}
which follows from the molecular structure of the proteins. Correspondingly,
the free $\rm Ca^{2+}$ efflux in Eq.~(\ref{eq meaneq}) gets the form

\begin{equation}
\label{eq efflux}
J_{\rm {e}}\left( c\right) =\rho _{\rm {p}}\, g_{\rm {p}}\left( c\right) \, 
\overline{I}_{\rm {p}}+\rho _{\rm {x}}\, g_{\rm {x}}\left( c\right) \, 
\overline{I}_{\rm {x}}\, .
\end{equation}

This is again formulated on the single protein level, where 
\( \rho _{\rm {p},\rm {x}} \)
represent the \emph{specific} surface densities of the membrane proteins
and \( \overline{I}_{\rm {p},\rm {x}} \) the \emph{universal} maximum
activity rates of the PMCA and the NCX, respectively. For the \emph{universal}
activity characteristics \( g_{\rm {p},\rm {x}} \) we assume Hill
equations

\begin{equation}
\label{eq. pumptauschprob}
g_{\rm {p}}\left( c\right) =\frac{\left( c\right) ^{n_{\rm {p}}}}
{\left( c\right) ^{n_{\rm {p}}}+\left( H_{\rm {p}}\right) ^{n_{\rm {p}}}}
\quad \textrm{and}\quad g_{\rm {x}}\left( c\right) =
\frac{\left( c\right)^{n_{\rm {x}}}}{\left( c\right) ^{n_{\rm {x}}}+
\left( H_{\rm {x}}\right) ^{n_{\rm {x}}}}
\end{equation}

which directly depend on the free $\rm Ca^{2+}$ concentration \( c \) \cite{Zador_etal90}.
\( H_{\rm {p},\rm {x}} \) are the half activation concentrations
and \( n_{\rm {p},\rm {x}} \) the Hill coefficients. This is justified
in the case of PMCA \cite{Elwess_etal97} while no explicit measurements
exist in the case of NCX. Nevertheless, it is a reasonable assumption
that the principal behavior is also well described by a Hill equation.

\subsection{Universal properties of the model}

The classification of the model parameters represents an important
element of our model because it enables us to separate those properties
which are basically independent of the cell-type from those which
are sensitive for differences between cell-types, compartments, and
experimental setups. The universal properties of the proteins are
postulated to remain invariant for all kinds of neurons and we extracted
these values from various single-protein experimental results. The
values of universal parameters for VDCC, PMCA and NCX as used in the
model are summarized in Tab.~\ref{tab parameter}.

\begin{table}[tb]
\renewcommand{\arraystretch}{1.1}
{\centering {\begin{tabular}{p{6cm}b{2cm}}\hline
value &	references\\\hline\hline
\multicolumn{2}{l}{\textbf{VDCC}}\\
\( \overline{g}_{\rm {v}}=14\, \rm {pS} \) & \cite{Fisher_etal90}\\
\( V_{\rm {h}}=-4\, \rm {mV} \) & \cite{Fisher_etal90}\\
\( \kappa =6.3\, \rm {mV} \) & \cite{Fisher_etal90}\\
\( \tau =1\, \rm {ms} \) & \cite{Magee_Johnston95}\\
\( \overline{V}_{\rm {eq}}=47\, \rm {mV} \) & \cite{Fisher_etal90}\\ \hline
\multicolumn{2}{l}{\textbf{PMCA}}\\
\( \overline{I}_{\rm {p}}=0.27 \times 10^{-20}\frac{\rm {C}}{\rm {ms}} \) & 
\cite{Elwess_etal97},\cite{Enyedi_etal93a}\\
\( n_{\rm {p}}=2 \) & \cite{Elwess_etal97}\\
\( H_{\rm {p}}=0.09\, \mu \rm {M} \) & \cite{Elwess_etal97}\\\hline
\multicolumn{2}{l}{\textbf{NCX}}\\
\( \overline{I}_{\rm {x}}=0.48 \times 10^{-18}\frac{\rm {C}}{\rm {ms}} \) &
\cite{Juhaszova_etal00}\\
\( n_{\rm {x}}=1 \) & see text\\
\( H_{\rm {x}}=1.8\, \mu \rm {M} \) & \cite{Blaustein_Lederer99}\\\hline
\end{tabular}}\par}
\caption{The values of universal model parameters. \label{tab parameter}}
\end{table}

\textbf{VDCC}. These channel-proteins are multi-subunit complexes which
form a voltage sensitive transmembrane pore. Six types of VDCC are
known, corresponding to the pharmacological properties (i.e. L-, N-,
P/Q-, R-type as high-voltage-activated (HVA) channels and T-type as
low-voltage-activated (LVA) channels. These types are primarily characterized
by genes encoding a different subunit (e.g. the \( \alpha _{1} \)
subunit) but an overall matching percentage of nearly \( 50\% \)
(for HVA channels only) \cite{Lacinova_etal00}. The $\rm Ca^{2+}$ influx
into the presynaptic terminal is dominated by the P/Q- and N-type
channels \cite{Koester_Sakmann00}. Each type of VDCC should have
the same biophysical properties in all kinds of tissues and animals.
Unfortunately the actual precision of the experiments, especially
taking into account the variety of experimental setups, does not allow
to quantitatively identify the biophysical properties corresponding
to each channel type \cite{Lorenzon_Foehring95,Tottene_etal96,Magee_Johnston95}.
The resulting uncertainties dominate the differences between the channel
types. Therefore, we chose the values from one single channel measurement
of N-type channels considering it as a representative member for the
VDCC. However, if more conclusive values for different VDCC-types
are found in future we may disentangle the influence of different
sub-types on $\rm Ca^{2+}$ transients. Concerning the measurements of the
P/Q- and N-type channels two additional problems exist. Most measurements
were performed with \( \rm {Ba}^{2+} \) as carrier ion and the obtained
conductivity differs from the $\rm Ca^{2+}$ conductivity \cite{Lorenzon_Foehring95}.
Additionally all measurements were made at room temperature 
(\( 20-25^{\circ }\rm {C} \))
and the biophysical properties are temperature dependent 
\cite{Coulter_etal89,Takahashi_etal91}.
Thus, the obtained $\rm Ca^{2+}$ influx through the VDCC differs between
measurements made at room temperature from those which were made at
blood temperature. We neglect both deviations because we could neither
find convincing extrapolations of the temperature behaviour nor a
sufficient relation of different charge carrier ions. Anyhow, those
deviations stay within the current experimental accuracy.

\textbf{PMCA}. There exist four different genes which encode the PMCA
(PMCA1-4) and all of them occur in different splicing variants
(labeled by small letters). These iso-forms have different kinetic
properties but again we assume that one iso-form should have the same
properties in each type of tissues and animal. The dominant iso-forms
of PMCA in the rat brain are the 1a-, 2a-, 2b-, 3a-, 3b-, and 4b-type
\cite{Penniston_Enyedi98,Filoteo_etal97,Lehotsky95}. Similarly to
the VDCC the measured kinetic properties vary considerably due to
the different experimental conditions for one type of PMCA 
\cite{Enyedi_etal91,Elwess_etal97,Filoteo_etal00}
and we chose as representative member the PMCA2a protein. The Hill
coefficient and the concentration of half activation are taken from
a measurement for rat PMCA \cite{Elwess_etal97} and the maximum pump
current is calculated from \cite{Elwess_etal97,Enyedi_etal93a}. We
neglect other regulation mechanisms of the PMCA-activity e.g. the
roles of calmodulin or of ATP (which is considered to be available
to a sufficient degree).

\textbf{NCX}. In the case of the NCX three different genes (NCX1-3)
with different splicing variants are known. Here, we do not specify
the type of NCX. We chose average values for the maximum current 
\cite{Juhaszova_etal00}
and for the concentration of half activation \cite{Blaustein_Lederer99},
and set the Hill coefficient of the NCX to one. It is worth emphasizing
that our results are not significantly altered for other Hill coefficient.
Other regulation mechanisms (e.g. the dependence on $\rm Na^{+}$) are not
considered in the model.

\subsection{Specific properties of the model\label{chapter surfdens}}

Per definition the specific parameters have to be adjusted to the
specific neuron type and compartment geometry in the experiment under
consideration. Especially, this concerns the rest state properties
\textbf{}(i.e. geometry factor \( G \), rest state membrane potential
\( V_{0} \), rest state intracellular \( c_{0} \) and extracellular
\( c_{\rm {ext}} \) $\rm Ca^{2+}$ concentration) which in general are indicated
in the experiments. This applies not necessarily to the dissociation
constant \( K_{\rm {m}} \) and the total concentration \( m_{0} \)
of the endogenous buffer. Here we chose Calmodulin as representative
member of buffer molecules and we assume that the four buffer binding
sites are independent. Our aim is to quantitatively compare the results
of our model with experiments. Because all visualizations of $\rm Ca^{2+}$
dynamics in neural compartments are performed with fluorescence indicators
we have to include the parameters specific for the used type of indicator.
The corresponding dissociation constant can be extracted from corresponding
data sheets and the indicator concentration is usually indicated in
the experimental protocol.

The most critical specific parameters are the membrane protein surface
densities (e.g. VDCC \( \rho _{\rm {v}} \), \textbf{}PMCA \( \rho _{\rm {p}} \)).
They are rarely known for specific cell types. Even more seldom
they are indicated in specific experiments. In addition measurements
of protein densities are usually valid for very specific areas on
the cell membrane. It is difficult to extrapolate those densities
to an average density on a whole presynapse as it is used in the model.
As our main interest points towards the analysis of $\rm Ca^{2+}$ transients
in response to various stimulation protocols we proceed as follows.
We use single action potential $\rm Ca^{2+}$ transients measured as fluorescence
signal to fit the unknown protein densities. Then the values found
are retained and used for the analysis of $\rm Ca^{2+}$ transients after
other stimulation protocols. This procedure is explained in more details
in section~\ref{chapter quamotest}.

\subsection{Equilibrium, stability and stimulation\label{chapter sss}}

To obtain a realistic simulation of $\rm Ca^{2+}$ dynamics the model has
to recover a stable equilibrium state after some stimulation. That
is achieved by the requirement that in the rest state of the neuron
(defined by the rest membrane potential \( V_{0} \) and the free
rest $\rm Ca^{2+}$ concentration \( c_{0} \)) the leak conductivity of
the membrane \( L \) exactly compensates the netto current densities
coming from VDCC (\( J_{\rm {i}} \)), PMCA and NCX (\( J_{\rm {e}} \)).

After application of any type of stimulation the free $\rm Ca^{2+}$ concentration
indeed returns to the rest concentration. We did not find any instabilities
of the $\rm Ca^{2+}$ dynamics within physiologically relevant stimulation
protocols. The model results are robust against variation of any parameters
on the qualitative level. Quantitatively the results are most sensitive
to variations of the membrane protein surface densities and of the
dissociation constant of the endogenous buffer.

For the stimulation protocols (single action potential and trains
of action potentials) we use membrane potentials \( V\left( t\right)  \)
which are simulated with a system of coupled differential equations
(Hodgkin-Huxley like, not shown here). The parameters of those equations have 
to be adjusted such that the stimulation shapes as found in the experiment under
consideration are appropriately reproduced.

\section{Results\label{chapter results}}

To check if our model correctly describes the presynaptic
$\rm Ca^{2+}$ dynamics in response to single action potentials, we tuned
the model with respect to an experiment measuring the $\rm Ca^{2+}$ dynamics
in presynaptic boutons of pyramidal neurons in the neocortex of rats
\cite{Koester_Sakmann00}. The corresponding $\rm Ca^{2+}$ transients can
be reproduced for reasonable membrane protein surface densities. Using
those parameters we can calculate the $\rm Ca^{2+}$ dynamics due to a \( 10\, \rm {Hz} \)
tetanus and compare the result to corresponding measurements \cite{Koester_Sakmann00}
(see section~\ref{chapter quamotest}). A qualitative discussion
of $\rm Ca^{2+}$ transients in response to various stimulation protocols
is provided in section~\ref{chapter ltestimprot}. Finally we will
quantitatively illustrate the influence of fluorescence indicators
on $\rm Ca^{2+}$ dynamics and discuss the implications for experiments using
fluorescence indicators to measure $\rm Ca^{2+}$ transients 
(see section~\ref{chapter fluoindltp}).

\subsection{Quantitative model test\label{chapter quamotest}}

At first we verify that our model can describe $\rm Ca^{2+}$ transients
evoked by single action potentials in single presynaptic boutons from
pyramidal neurons in the neocortex of rats on the basis of a very detailed
corresponding fluorescence measurements \cite{Koester_Sakmann00}.
The universal parameters should especially apply for this system
and are used as shown in Tab.~\ref{tab parameter}. The specific
parameters are determined directly and indirectly by the measurement
\cite{Koester_Sakmann00} and by using independent experimental results.
We calculated the geometry factor with the help of a fluorescence
figure of the whole bouton (Fig.~3A in \cite{Koester_Sakmann00}).
We determined the dissociation constant of the endogenous buffer \( K_{\rm {m}} \)
using the observation that approximately only \( 1\% \) of the total
incoming intracellular $\rm Ca^{2+}$ remains free \cite{Koester_Sakmann00}.
The surface density of the NCX \( \rho _{\rm {x}} \) was determined
with the help of a measurement which investigates its ratio to the
PMCA density \( \rho _{\rm {p}} \) \cite{Juhaszova_etal00}. The
values for the specific parameters are summarized in Tab.~\ref{tab specpara}.
They have to be carefully interpreted as an approximation for the
neuron type under consideration. Please note that only two
parameters remained free to adjust the model to the experimental
data: The densities of the PMCA and of the VDCC. Note, that any dependence of channel
densities on intracellular $\rm Ca^{2+}$ \cite{Siegel_etal94}  has not been considered here.

\begin{table}[tb]
\renewcommand{\arraystretch}{1.1}
{\centering {\begin{tabular}{m{6cm}b{2cm}} \hline
value & references\\\hline\hline
\multicolumn{2}{l}{\textbf{geometry (steady state)}}\\
\( G=3/0.5\, \mu \rm {m} \) & \cite{Koester_Sakmann00}\\
\( V_{0}=-70\, \rm {mV} \) & \cite{Koester_Sakmann00}\\
\( c_{0}=0.1\, \mu \rm {M} \) & \cite{Hille92}\\
\( c_{\rm {ext}}=1.5\, \rm {mM} \) & \cite{Hille92}\\\hline
\multicolumn{2}{l}{\textbf{protein densities (single action potential)}}\\
\( \rho _{\rm {v}}=3.1/\mu \rm {m}^{2} \) & adjusted\\
\( \rho _{\rm {p}}=9200/\mu \rm {m}^{2} \) & adjusted\\
\( \rho _{\rm {x}}=0.033*\rho _{\rm {p}} \) & \cite{Juhaszova_etal00}\\\hline
\multicolumn{2}{l}{\textbf{protein densities (\( 10\, \rm {Hz} \) tetanus)}}\\
\( \rho _{\rm {v}}=3.7/\mu \rm {m}^{2} \) & adjusted\\
\( \rho _{\rm {p}}=8300/\mu \rm {m}^{2} \) & adjusted\\
\( \rho _{\rm {x}}=0.033*\rho _{\rm {p}} \) & \cite{Juhaszova_etal00}\\\hline
\multicolumn{2}{l}{\textbf{endogenous buffer}}\\
\( K_{\rm {en}}=0.5\, \mu \rm {M} \) & \cite{Koester_Sakmann00}\\
\( b_{\rm {en}}^{0}=4*30\, \mu \rm {M} \) & \cite{Schiegg_etal95},\cite{Carafoli87}\\\hline
\multicolumn{2}{l}{\textbf{fluorescence indicator}}\\
\( K_{\rm {ex}}=6\, \mu \rm {M} \) & \cite{Koester_Sakmann00}\\
\( b^{0}_{\rm {ex}}=100,500\, \mu \rm {M} \) & \cite{Koester_Sakmann00}\\
\( \left( \Delta f/f\right) ^{100\mu \rm {M}}_{\rm {max}}=1.5 \) & \cite{Koester_Sakmann00}\\
\( \left( \Delta f/f\right) ^{500\mu \rm {M}}_{\rm {max}}=2.3 \) & adjusted\\\hline
\end{tabular}}\par}
\caption{The values of specific model parameter. \label{tab specpara}}
\end{table}

In general $\rm Ca^{2+}$ measurements in neurons use fluorescence indicators
to visualize the $\rm Ca^{2+}$ concentration and distribution. As already
pointed out such fluorescence indicators act as an exogenous buffer
system which, therefore, has been included into the model prescription
(see section~\ref{chapter buffering}). With the help of a \( 1:1 \)
complexation ansatz \cite{Koester_Sakmann00,Jaffe_etal94,Grynkiewicz_etal85}
we translated the simulation of the $\rm Ca^{2+}$ dynamics into a simulation
of the fluorescence signal

\begin{equation}
\label{eq fluo}
\Delta f/f=\left( \Delta f/f\right) _{\rm {max}\, }\frac{c\left( t\right) -
c_{0}}{c\left( t\right) +K_{\rm {ex}}}\, \, .
\end{equation}

\( \Delta f/f \) is the relative fluorescence change, 
\( \left( \Delta f/f\right) _{\rm {max}} \) represents its maximum, and 
\( K_{\rm {ex}} \) is the dissociation constant of the indicator. The properties 
of the used indicators (Magnesium Green (Molecular Probes)) are shown in 
Tab.~\ref{tab specpara}. Note, that in the experiment different concentrations of 
indicator have been used \cite{Koester_Sakmann00}.

\begin{figure}[tb]
{\centering \resizebox*{8cm}{!}{\includegraphics{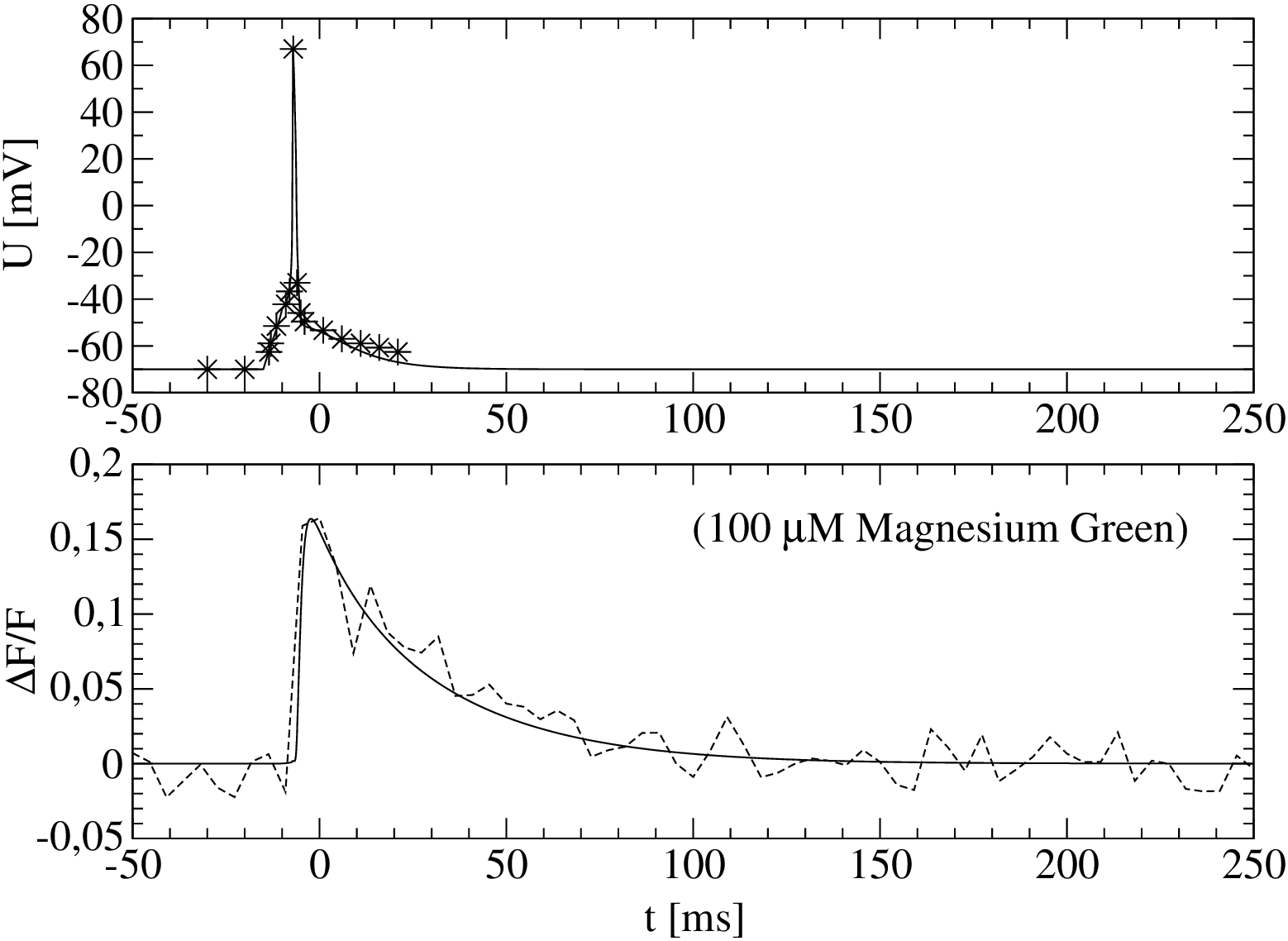}} \par}
\caption{The $\rm Ca^{2+}$ transient (shown
as relative fluorescence change with \( 100\, \mu \rm {M} \) Magnesium
Green) evoked by single action potentials in single
boutons of pyramidal neurons in the rat neocortex. The upper panel
shows the simulated action potential compared to the one used in experiment
(stars). The lower panel compare the experimental results (with kind
permission of H.J. Koester and B. Sakmann \cite{Koester_Sakmann00},
Fig.~15A) to the model results. The parameters as listed in tables
Tab.~\ref{tab parameter} and Tab.~\ref{tab specpara} have been
used.\label{pic sakfluosap}}
\end{figure}

The only remaining free parameters are the surface densities of the
VDCC \( \rho _{\rm {v}} \) and the PMCA \( \rho _{\rm {p}} \). We
simulate the amplitude and the shape of the action potential as applied
to the presynapse in the experiment and fit the protein densities
(see Tab.~\ref{tab specpara}) such that the simulation result agrees
with the measured fluorescence signal evoked by a single action potential
(Fig.~15C in \cite{Koester_Sakmann00}). Note that the experiment
has been executed with \( 100\, \mu \rm {M} \) Magnesium Green and 
\( \left( \Delta f/f\right) ^{100\, \mu \rm {M}}_{\rm {max}} \)
has to be chosen correspondingly (see Tab.~\ref{tab specpara}).
The result is shown in Fig.~\ref{pic sakfluosap}. The amplitude
and the shape of the simulated fluorescence response are in good agreement
with the data observed in experiment.

\begin{figure}[tb]
{\centering \resizebox*{8cm}{!}{\includegraphics{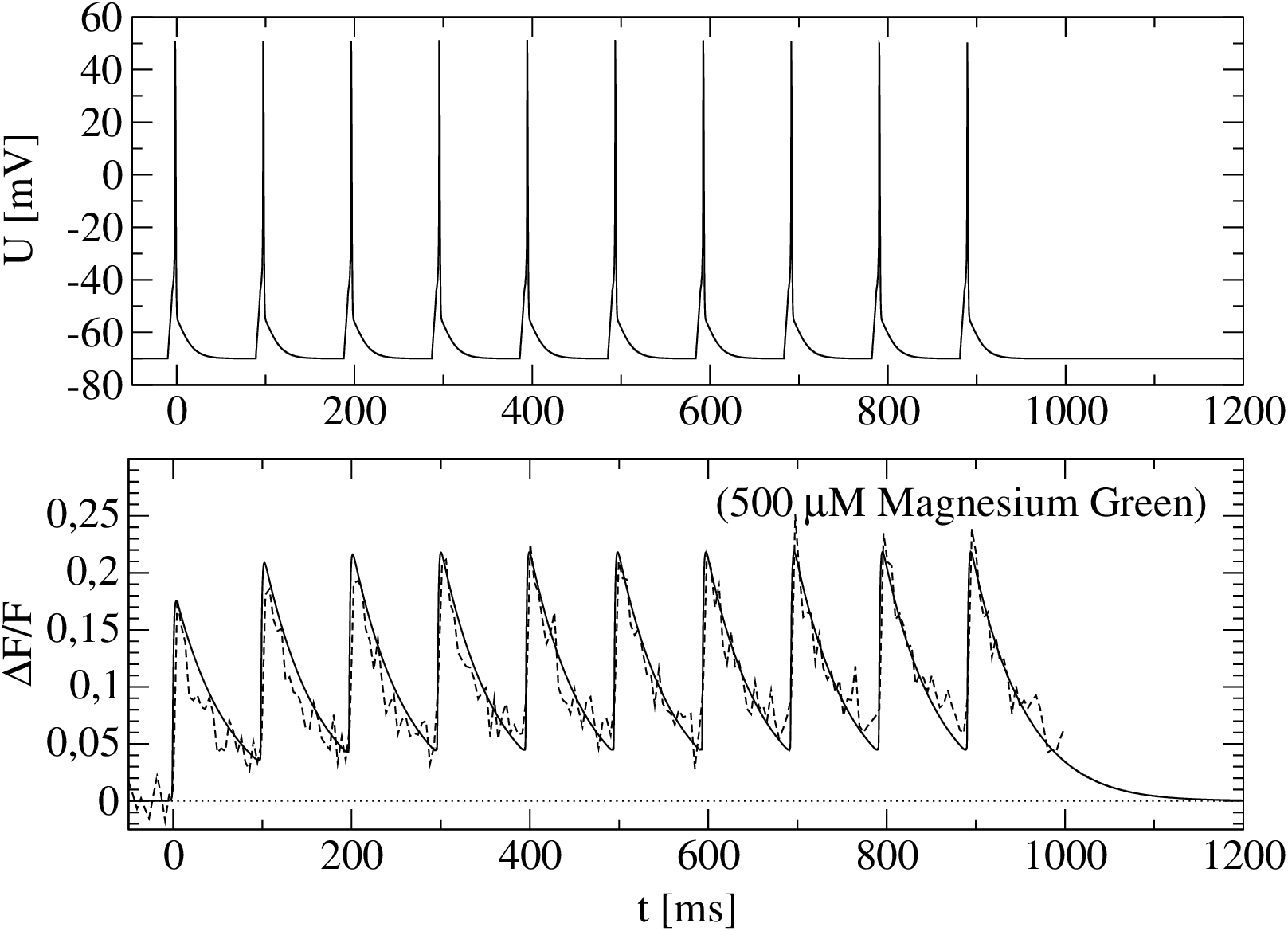}} \par}
\caption{The $\rm Ca^{2+}$ transients (shown as relative fluorescence change
with \( 500\, \mu \rm {M} \) Magnesium Green) evoked
by a \( 10\, \rm {Hz} \) tetanus in single boutons of pyramidal neurons
in the rat neocortex. The measured $\rm Ca^{2+}$ signal observed in experiment
(with kind permission of H. J. Koester and B. Sakmann
\cite{Koester_Sakmann00}, Fig.~9) is compared to the corresponding
model results (lower panel). The upper panel shows the simulated stimulation
protocol. The parameters as listed in Tab.~\ref{tab parameter} and Tab.~\ref{tab
specpara} have been used.
\label{pic sakfluo10hz}}
\end{figure}

The next step is to investigate whether the same model without changing
the model assumptions is able to describe $\rm Ca^{2+}$ transients evoked
by more complex stimulation protocols. We compare the fluorescence
response evoked by a \( 10\, \rm {Hz} \) tetanus to the model predictions.
To this end the action potentials in the model have to be adjusted
to those used in experiment (Fig.~9; upper panel in \cite{Koester_Sakmann00}).
The experiment has been performed with \( 500\, \mu \rm {M} \) Magnesium
Green and we corrected the indicator concentration
correspondingly. The maximal relative fluorescence change
\( \left( \Delta f/f\right) ^{500\, \mu \rm {M}}_{\rm {max}} \)
has not been stated explicitly. Therefore, we determine this value
by the requirement that the $\rm Ca^{2+}$ signal in response to a single
action potential (Fig.~15C; middle column in \cite{Koester_Sakmann00})
is reproduced correctly (data not shown). Finally, the protein surface
densities are adjusted to reproduce the $\rm Ca^{2+}$ response to the first
action potential of the tetanus (see Tab.~\ref{tab specpara}). The
subsequent $\rm Ca^{2+}$ signal as predicted by the model turns out to be
in perfect agreement with the fluorescence signal (see Fig.~\ref{pic sakfluo10hz}).
Note, that all others parameters of the model remained unchanged.
We conclude that the model once fitted to the $\rm Ca^{2+}$ signal in response
to single action potentials in a specific type of neuron describes
the $\rm Ca^{2+}$ signal due to tetanus stimulation without further changes.

\subsection{$\rm Ca^{2+}$ dynamics in response to different stimulation protocols
\label{chapter ltestimprot}}

\begin{figure}[tb]
{\centering \resizebox*{8cm}{!}{\includegraphics{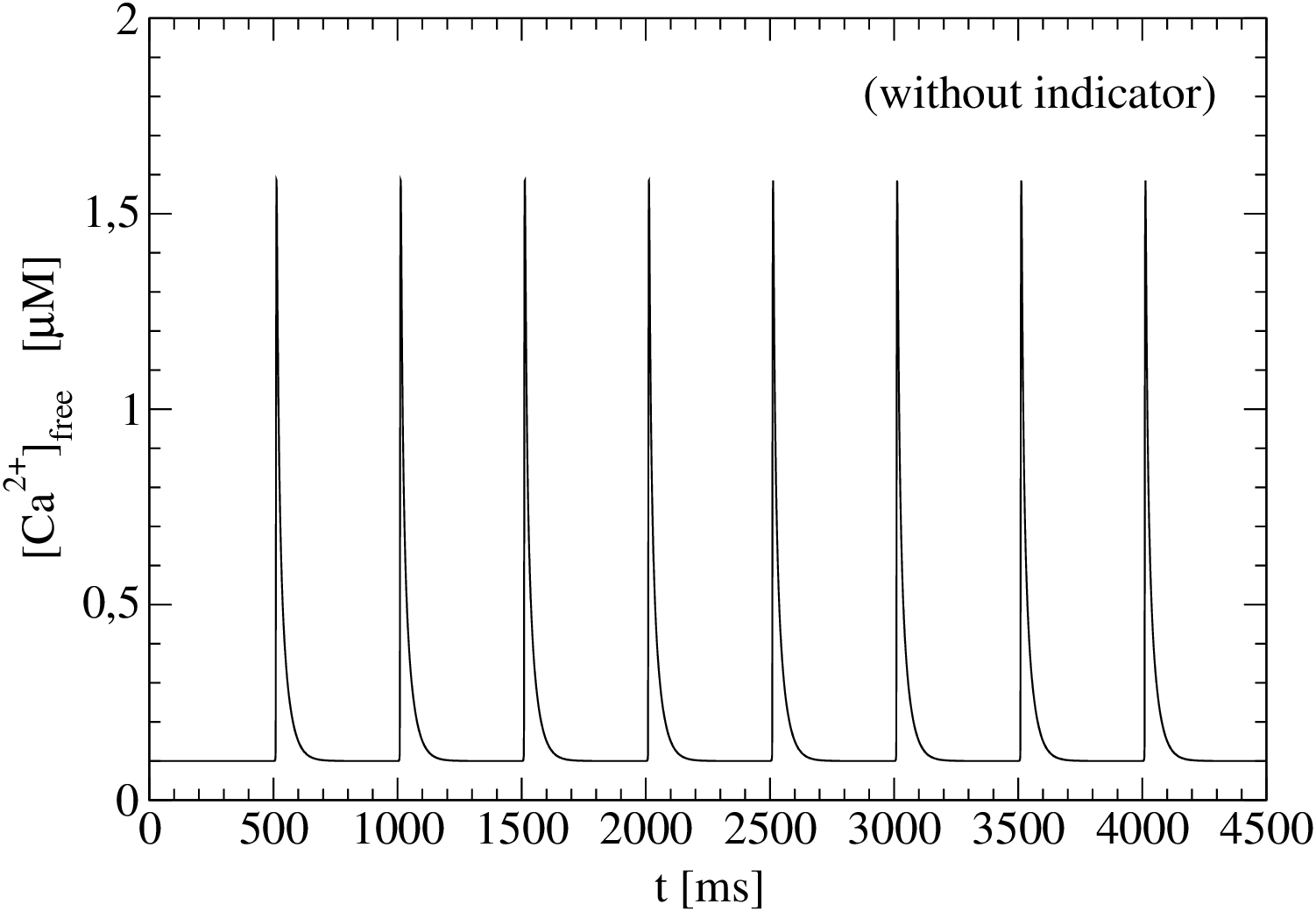}} \par}
\caption{The $\rm Ca^{2+}$ transients (without
indicator) evoked by a \( 2\, \rm {Hz},3.5\, \rm {s} \) tetanus. 
All other parameters as in Tab.~\ref{tab parameter}
and Tab.~\ref{tab specpara}. The $\rm Ca^{2+}$ response to each action
potential stimulation remain independent from each other.
\label{pic ltdtet}}
\end{figure}

The main question we address is whether
there are characteristic differences between the different 
$\rm Ca^{2+}$ transients or not. In the low frequency domain 
(\( 2\, \rm {Hz},3.5\, \rm {s} \)) the $\rm Ca^{2+}$ response 
appears as a train of independent single action
potential responses (Fig.~\ref{pic ltdtet}). The $\rm Ca^{2+}$ concentration
follows the activation by each action potential. Shape and amplitude
of each $\rm Ca^{2+}$ response is unaltered compared to the response to
a single action potential. This frequencies range is typically used for LTD induction
\cite{Connor_etal99}.

For higher frequencies (\( 20\, \rm {Hz},350\, \rm {ms} \)) a new baseline 
in the $\rm Ca^{2+}$ concentration arises during stimulation (Fig.~\ref{pic
transtet}). The $\rm Ca^{2+}$ pumps and the $\rm Na^{+}$/$\rm Ca^{2+}$ 
exchangers have not enough time to return the neuron into its rest state. 
This implies an overlap of $\rm Ca^{2+}$ response to subsequent action potentials. 
Note, that this qualitative behavior has been observed for example in experiments performed 
on dendritic spines of pyramidal neurons \cite{Helmchen_etal96}. However, we can
not expect that the presynaptic model also quantitatively describes
the $\rm Ca^{2+}$ response in dendritic spines correctly (especially without including diffusion
in the model).

\begin{figure}[tb]
{\centering \resizebox*{8cm}{!}{\includegraphics{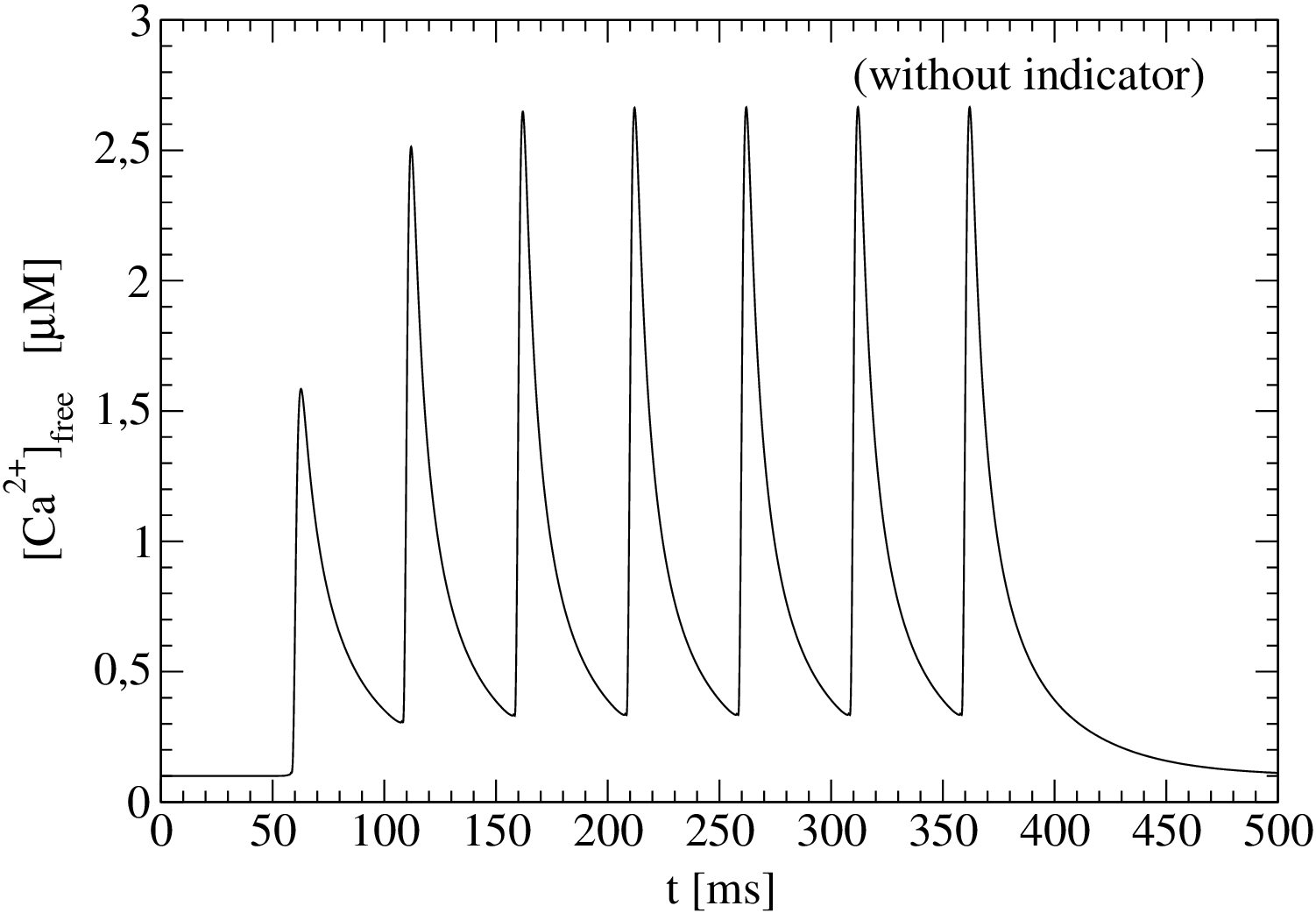}} \par}
\caption{The $\rm Ca^{2+}$ transients (without
indicator) evoked by a \( 20\, \rm {Hz},350\, \rm {ms} \) tetanus. 
All other parameters as in Tab.~\ref{tab parameter} and Tab.~\ref{tab specpara}. A
new baseline of the $\rm Ca^{2+}$ concentration arises during stimulation.
\label{pic transtet}}
\end{figure}

This strong increase of the peak and baseline and especially of the average
$\rm Ca^{2+}$ concentration for stimulations with higher frequencies is a 
significant difference compared to the $\rm Ca^{2+}$ response to a low-frequency 
tetanus. One may suspect that this behaviour is an important part of a process 
to change synaptic efficiency. For example an accumulation of residual free 
$\rm Ca^{2+}$ is responsible for synaptic facilitation in presynaptic boutons 
\cite{Fogelson_Zucker85,Charlton_etal82}. Also for the induction of LTP it is well 
established fact, that the intensity of stimulation protocols have to overcome some
threshold \cite{Connor_etal99}. The strong increase of the $\rm Ca^{2+}$ 
concentration emerges for frequencies that are in the same range as this threshold 
\cite{Rick_Milgram96,Gamble_Koch87}. This hypothesis also agrees with the 
experimental fact that a modest $\rm Ca^{2+}$ influx causes LTD, whereas a large 
$\rm Ca^{2+}$ influx triggers LTP \cite{Connor_etal99,Yang_etal99}. In general the 
role of enhanced average $\rm Ca^{2+}$ concentrations for changes of synaptic
efficiency has been pointed out by \cite{Wang_etal96}. In order to look for a
specific key of a concrete induction process of one of the different forms of 
synaptic plasticity one has to consider the synapse in more detail. For instance 
exocytosis steps \cite{Rosenmund_etal02}, buffer properties \cite{Tang_etal00}, 
diffusion effects \cite{Zucker_Stockbridge83} and protein kineses
\cite{Malenka_etal89} are important elements of the induction machinery. This is a 
task for future modeling work.

\begin{figure}[tb]
{\centering \resizebox*{8cm}{!}{\includegraphics{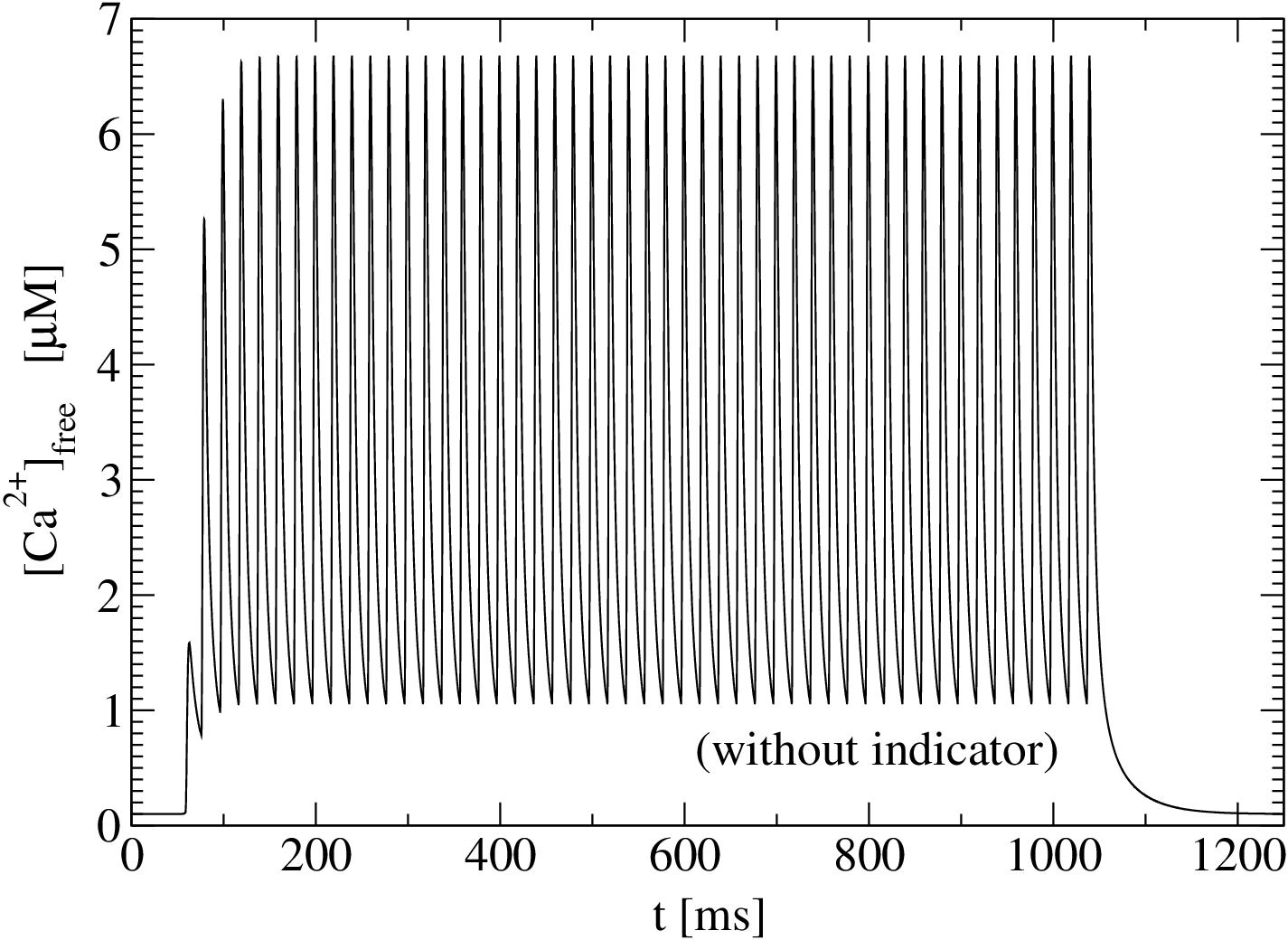}} \par}
\caption{The $\rm Ca^{2+}$ transients (without
indicator) evoked by a \( 50\, \rm {Hz},1\, \rm {s} \) tetanus. 
All other parameters as in Tab.~\ref{tab parameter}
and Tab.~\ref{tab specpara}. There is a more pronounced enhancement
of the $\rm Ca^{2+}$ baseline.\label{pic ltptet}}
\end{figure}

During the application of a high-frequency 
tetanus (\( 50\, \rm {Hz},1\, \rm {s} \))
the induced $\rm Ca^{2+}$ spikes remain in phase with the stimulating membrane
potential. The enhancement of the $\rm Ca^{2+}$ baseline becomes even more
pronounced (Fig.~\ref{pic ltptet}). Again the new baseline saturates
during stimulation but on a higher level compared to the \( 20\, \rm {Hz} \)
stimulation. The exact shape of the $\rm Ca^{2+}$ signal is a result of
an interplay of membrane proteins and $\rm Ca^{2+}$ buffers. However, we
observed some relations between the general appearance of the $\rm Ca^{2+}$
signal and some specific neuron properties. The surface density of
VDCC basically determines the amplitude of the $\rm Ca^{2+}$ spike on the
top of the baseline. The dissociation constant determines the height of the new 
baseline. The time scale on which the $\rm Ca^{2+}$ concentration returns to its rest
state is governed by the surface densities of the PMCA, NCX and the dissociation constant
of the buffer. We emphasize that even higher stimulation frequencies lead to still 
higher baseline levels of the $\rm Ca^{2+}$ concentration. The reached baseline grows
quadratically with the stimulation frequency up to \( 100\, \rm {Hz} \).

\subsection{Fluorescence indicators disturb the intrinsic $\rm Ca^{2+}$ dynamics
\label{chapter fluoindltp}}

\begin{figure}[tb]
{\centering \resizebox*{8cm}{!}{\includegraphics{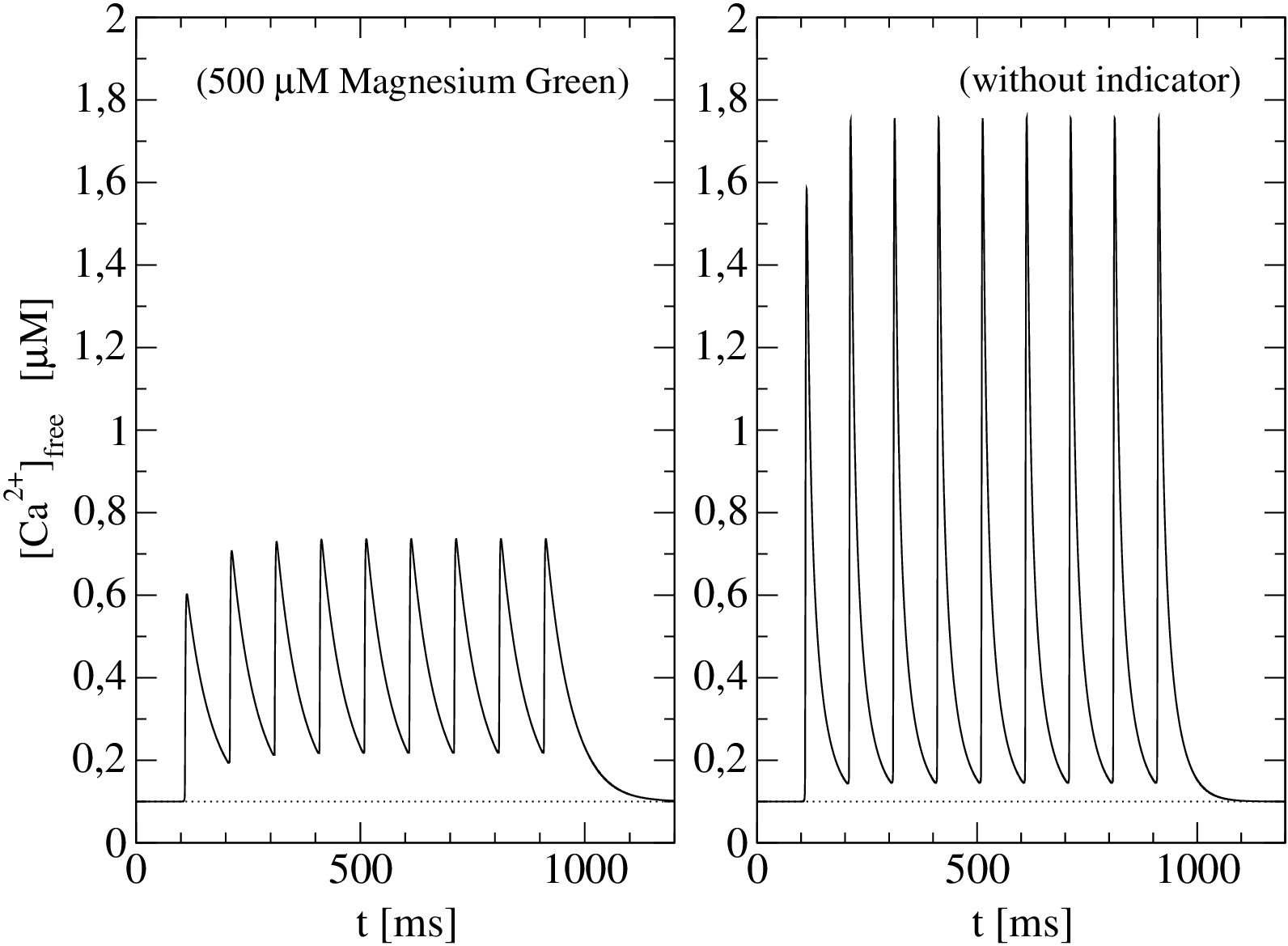}} \par}
\caption{$\rm Ca^{2+}$ transients evoked
by a single action potential. The upper panel shows the $\rm Ca^{2+}$ transient
for a neuron with \( 100\, \mu \rm {M} \) Magnesium Green, and the lower panel the 
$\rm Ca^{2+}$ transient which would occur
without indicator. The indicator changes the amplitude and the shape
of the $\rm Ca^{2+}$ transient considerably.\label{pic withwithoutsap}}
\end{figure}

In section~\ref{chapter ltestimprot} we have seen that for stimulations
with frequencies above some threshold frequency a new baseline of
the $\rm Ca^{2+}$ concentration arises. In the following we show that the
use of fluorescence indicators for the visualization of $\rm Ca^{2+}$ in
experiments alters the $\rm Ca^{2+}$ response and especially the emergence
of a new baseline considerably. In Fig.~\ref{pic withwithoutsap}
the $\rm Ca^{2+}$ transients evoked by single action potentials with and
without fluorescence indicator is compared (\( 100\, \mu \rm {M} \)
Magnesium Green). The amplitude of the $\rm Ca^{2+}$
signal is decreased if the fluorescence indicator is used. In addition
the $\rm Ca^{2+}$ relaxation time (needed to recover the rest $\rm Ca^{2+}$ concentration)
becomes considerably larger using the fluorescence indicator. These
relations are not surprising as the indicator binds an important part
of incoming free $\rm Ca^{2+}$ ions and reduces the free $\rm Ca^{2+}$ concentration.
Note, that the larger relaxation times are not caused by a decreased pump activity
which is induced by the lower $\rm Ca^{2+}$ peak. This has consequences for the 
interpretation of experiments using fluorescence indicators. 

\begin{figure}[tb]
{\centering \resizebox*{8cm}{!}{\includegraphics{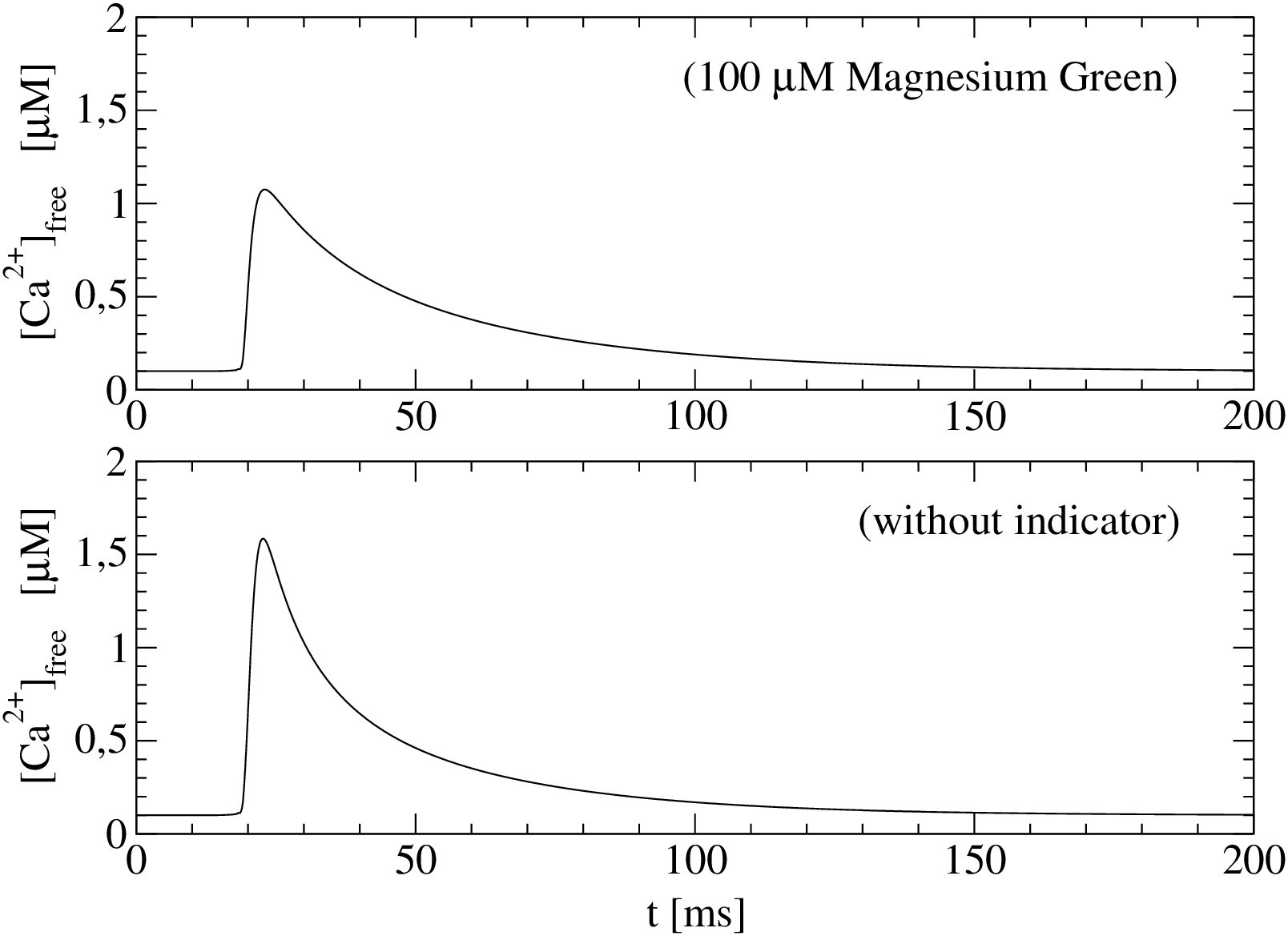}} \par}
\caption{$\rm Ca^{2+}$ transients evoked
by a \( 10\, \rm {Hz} \) tetanus. The left panel shows the $\rm Ca^{2+}$
transient with \( 500\, \mu \rm {M} \) Magnesium Green. The right panel shows the 
$\rm Ca^{2+}$ transient without indicator.
All other parameters as in Tab.~\ref{tab parameter} and Tab.~\ref{tab specpara}.
The presence of an indicator changes the $\rm Ca^{2+}$ signal and the height
of the emerging baseline considerably.\label{pic withwithout10hz}}
\end{figure}

The fact that fluorescence indicators act as an additional 
intracellular buffer changing the shape of the calcium transient
has been pointed out before \cite{Smith_etal96a,Matveev_etal02}.
In Fig.~\ref{pic withwithout10hz}
the influence of the fluorescence indicator (\( 500\, \mu \rm {M} \)
Magnesium Green) on the $\rm Ca^{2+}$ transients evoked
by a \( 10\, \rm {Hz} \) tetanus in a pyramidal neuron of the rat
neocortex \cite{Koester_Sakmann00} is shown. The left panel (with
fluorescence indicator) clearly exhibits a new baseline of the $\rm Ca^{2+}$
concentration whereas the baseline remains practically unchanged in
the right panel (without fluorescence indicator). Because of the longer
relaxation time for the $\rm Ca^{2+}$ spikes the fluorescence indicator
facilitates the emergence of a new baseline. Using a fluorescence
indicator the $\rm Ca^{2+}$ spikes overlap already for lower frequencies.
Consequently, the threshold stimulation frequency for the emergence
of a new baseline is higher for neurons without indicator compared
to neurons that has been treated with indicator. A interesting conclusion from
that fact is that the threshold frequency for the appearance of enhanced $\rm Ca^{2+}$ 
level changes to lower frequencies if a buffer is present. This may have 
implications for the induction of LTP in experiments using fluorescence indicators. 
Note, that the right panel in Fig.~\ref{pic withwithout10hz} provides a quantitative
prediction of the $\rm Ca^{2+}$ signal in the pyramidal neuron of the neocortex
as it would have been without the use of $\rm Ca^{2+}$ indicator starting
from the $\rm Ca^{2+}$ measurement \cite{Koester_Sakmann00}. 
The same reconstruction procedure may be applied to other experiments in future.

\section{Discussion\label{chapter discussion}}

We developed a model for the presynaptic $\rm Ca^{2+}$ dynamics which
includes VDCC, PMCA, and NCX, as well as endogenous buffers and fluorescence
indicators. Those ingredients turned out to be sufficient to quantitatively reproduce
a fluorescence measurement of intracellular free $\rm Ca^{2+}$ transients
in response to single action potentials in pyramidal neurons of the rat neocortex. 
On the basis of the parameter set determined by the single action potential 
$\rm Ca^{2+}$ transients the model predicts the $\rm Ca^{2+}$ response to 
\( 10\, \rm {Hz} \) tetanus (using the same neuron type) in a quantitatively accurate
way. We conclude that from the point of view of the model $\rm Ca^{2+}$ induced 
$\rm Ca^{2+}$-release is not necessarily involved into the presynaptic $\rm Ca^{2+}$ 
dynamics at least for this specific type of neuron.

We would like to emphasize that the single action potential $\rm Ca^{2+}$
response has been produced by fitting only two parameters, i.e. the
average membrane protein surface densities of the VDCC and the PMCA.
All other parameters have been determined either by the experiment
itself or by independent experiments. The resulting values for the
VDCC density are in good agreement with experimental measured values
\cite{Magee_Johnston95}. The PMCA density in contrast turn out to
be rather high but remain within a range
of densities that has been observed in experiment \cite{Khananshvili98}.
One may think of inverting the line of argumentation and interprete
the resulting protein densities as prediction for an average density
on the whole bouton. In view of the difficulties to measure such densities
this provides an attractive possibility.

The classification of parameters into universal and specific ones
opens the possibility to adjust the model to other experiments without
altering the model pillars. The (universal) properties of the VDCC
for example have been determined using corresponding single channel
experiments. In this context one may think of the necessity to include
more than one HVA channel type into the model instead of restricting
oneself to one representative type. However, a corresponding analysis
revealed that the resulting $\rm Ca^{2+}$ transients remain basically unchanged
using two different HVA channels. We conclude that the measurement
of $\rm Ca^{2+}$ transients does not allow to distinguish between different
HVA channel subtypes. This may be different if considering LVA channels
in the model.

We qualitatively investigated the general behaviour of $\rm Ca^{2+}$ transients
in response to different stimulation protocols. For \( 2\, \rm {Hz} \) 
tetanus we found no characteristic feature of the $\rm Ca^{2+}$ signal.
This agrees with the interpretation, that synaptic facilitation
is triggered by calcium enhancements, while synaptic depression is a
consequence of reduced release site activity \cite{Dittman_etal00}. 
We found a characteristic difference between $\rm Ca^{2+}$
transients in response to  \( 2\, \rm {Hz} \) and \( 50\, \rm {Hz} \)
stimulations. At some threshold stimulation frequency the $\rm Ca^{2+}$
signal does not return to its rest state level after each spike. Instead
a new baseline emerges on a higher level and the $\rm Ca^{2+}$ spikes develop
on top of this baseline. This general behaviour has also been found in
postsynaptic experiments before \cite{Helmchen_etal96}. 
Indeed, the strong increase in the $\rm Ca^{2+}$ signal
for above threshold stimulation frequencies 
can be interpreted to correspond to 
the stimulation threshold for changing synaptic efficiency \cite{Rozov_etal01}.
In other words, the emergence of an enhanced
baseline and, as a consequence, of an enhanced average value
and increased peak values of the $\rm Ca^{2+}$
concentration may be thought of as a necessary requirement for
induction of changes in the synaptic efficiency. The height
of the baseline increases quadratically with increasing stimulation
frequencies up to frequencies for which the buffer becomes saturated
with $\rm Ca^{2+}$. For higher frequencies the baseline increases only
linearly.

Already on the level of calcium transients averaged over the whole bouton we can
confirm the result, that the endogenous and exogenous buffer concentrations
as well as their saturation properties crucially shift the threshold
of the stimulation frequency  \cite{Rozov_etal01,Smith_etal96a,Matveev_etal02}. 
Especially, it is important to realize the strong influence of indicators on the
intracellular free $\rm Ca^{2+}$ signal \cite{Hansel_etal96,Hansel_etal97,Williams_Johnston89}.
Therefore, we investigated this problem in the framework of the model in more details.
On the level of single action potentials the use of indicator leads
to considerably smaller $\rm Ca^{2+}$ spikes which relaxed more slowly to
the rest concentration. This implies that the use of indicator shifts
the above mentioned threshold stimulation frequency. Therefore,
a quantitative evaluation of fluorescence measurements should always
include the effect of the indicator \cite{Nowycky_Pinter93,Tank_etal95}.
The model presented here, provides a tool to reconstruct the 
averaged $\rm Ca^{2+}$
transient as it would have bean without indicator in fluorescence
measurements. This has been done in this article for the experiment
of \cite{Koester_Sakmann00}. It would be interesting in future to
make use of the parameter classification, to enlarge the model to
calcium and buffer diffusion, and to apply the model to
other experiments, especially when using other neuron types. A comparative
analysis of presynaptic $\rm Ca^{2+}$ transients in different neurons may
reveal neuron-type specific characteristics of the $\rm Ca^{2+}$
dynamics.

\subsection*{Acknowledgments}

We are indebted to H. J. Koester and B. Sakmann for providing us with
the fluorescence data \cite{Koester_Sakmann00}.


\bibliographystyle{elsart-num}
\bibliography{ErlerMeyerhermannSoff03}


\end{document}